# Observation of intrinsic size effects in the optical response of individual gold nanoparticles

*Stéphane Berciaud, Laurent Cognet, Philippe Tamarat and Brahim Lounis**

Centre de Physique Moléculaire Optique et Hertzienne, CNRS (UMR 5798) et Université Bordeaux I,

351, cours de la Libération, 33405 Talence Cedex, France

AUTHOR EMAIL ADDRESS: b.lounis@cpmoh.u-bordeaux1.fr

ABSTRACT

The Photothermal Heterodyne Imaging method is used to study for the first time the absorption spectra of individual gold nanoparticles with diameters down to 5 nm. Intrinsic size effects which result in a broadening of the Surface Plasmon Resonance are unambiguously observed. Dispersions in the peak energies and homogeneous widths of the single particle resonances are revealed. The experimental results are analyzed within the frame of Mie theory.

Noble metal nanoparticles are strong absorbers and scatterers of visible light due to Surface Plasmon Resonance (SPR). The corresponding resonant peak energies and linewidths are sensitive to the nanoparticle size, shape and nano-environment[1]. Moreover, plasmon resonances allow for the enhancement and manipulation of local electromagnetic fields at nanoparticle surfaces[2-4]. Because of these singular optical features, noble metal nanoparticles stimulate great interests for implementations in photonics[5] or biotechnology[6,7].

The spectral properties of the SPR have been extensively studied on ensembles of noble nanoparticles and compared to Mie theory[8,9]. For rather large nanoparticles (diameter $D \geq 20$nm), the resonant peak energy $E_R$ experiences a red-shift with increasing sizes, due to retardation effects as well as to increasing contributions from multipolar terms. In addition, for nanoparticles larger than 50 *nm*, radiative damping of the collective electronic excitation significantly broadens the linewidth of the SPR. For smaller nanoparticles ($D < 20$ *nm*), these extrinsic size-effects become negligible and intrinsic size-effects prevail. They account for size-dependent modifications of the dielectric constant with respect to the bulk values, due to additional surface damping[1,10] and translate into a broadening of SPR linewidths with decreasing sizes, and to a lower extent, into slight shifts of $E_R$. Comparisons between theory and ensemble measurements are still a matter of debate because the inhomogeneities in nanoparticle size, shape and local environment experimentally hide the homogenous width of the SPR.

For large particles, scattering-based methods allowed to perform SPR measurements on individual particles and to study the extrinsic size-effects[11,12]. No intrinsic-size effects from additional surface damping were observed since a good agreement was found with Mie theory using the bulk values for the metal dielectric function[13]. Because the scattering cross-section of nanoparticles decreases as $D^6$, these methods are limited to nanoparticles with $D \geq 20$ *nm*. For smaller particles, different methods have been proposed to access the homogeneous SPR linewidth. Persistent spectral hole-burning techniques have recently been applied[14]: by the extrapolation of measurements performed at high intensities, homogeneous width values of a subpopulation of metal nanoparticles with a given resonant energy could be extracted from the inhomogeneous band. Concurrently, it was recently shown that plasmon resonance of single nanoparticles smaller than 10 *nm* can be measured by detecting the scattered field by one particle, through an interferometric method[15]. In that case, the measured signal scales as $D^3$ similarly to the absorption cross-section[16].

In this letter, we used a recently developed photothermal far-field absorption-based method[17], which allows to perform for the first time a systematic study of absorption spectra of individual gold nanoparticles with diameters down to 5 *nm* with a high signal-to-noise ratio. We found that a very small ellipticity in the nanoparticle shape results in a significant dispersion on the peak resonant energy $E_R$. We directly measured the homogeneous SPR linewidths and show that for nanoparticles with $D \leq 10 nm$ they are significantly affected by surface damping mechanisms leading to reduced SPR decoherence times.

We applied the Photothermal Heterodyne Imaging (PHI) method which allows for the detection of very small gold nanoparticles down to 1.4 *nm* in diameter[17]. It uses a combination of two laser beams: a intensity-modulated heating beam, close-to-resonance, and a cw off-resonance probe beam. Absorption of the heating beam by a nanoparticle induces a time modulated increase of the temperature in the vicinity of the nanoparticle. Propagation of the probe beam through the resulting time-modulated index of refraction profile, produces a frequency shifted scattered field which is detected by its beatnote with the probe field on a fast photodiode. The signal is thus directly proportional to the absorption cross section $\sigma_{abs}$ of the nanoparticle. The experimental setup is presented in Figure 1a. The heating beam was delivered by a tunable cw dye laser (Coherent 599, tuning range 515-580 *nm* i.e. photon energy range 2.41-2.14 *eV*). The intensity modulation of the beam (*300 kHz*) was performed by an acousto-optic modulator (AOM). Wavelength-dependent angular deviations were compensated by double pass through the AOM. The beam was linearly polarized and overlaid with a probe beam (HeNe laser, 633 *nm*) and focused on the sample.

In order to avoid any dependence of the relaxation dynamics of electrons on the absorbed power, we performed the experiments in the weak excitation regime where the average time between two successive absorption events is significantly longer than the relaxation times in metal nanoparticles (of the order of few *ps*)[18]. This corresponds to heating powers ranging from ≈ 1 *µW* to ≈ 500 *µW* (for $D$ = 33 *nm* to $D$ = 5 *nm* respectively), allowing to image the individual nanoparticles with signal-to-noise ratios greater than 10 at the peak resonant energies for 10ms integration times (see Figure 1b and c).

We prepared the samples by spin-coating solutions of gold nanoparticles (average diameter: 5 *nm*, 10 *nm*, 20 *nm*, 33 *nm*, diluted in 2% mass polyvinyl-alcohol (PVOH) aqueous solution) onto clean microscope coverslips. The dilution and spinning rates were chosen such that the final density of nanoparticles in the samples was less than 1 $µm^{-2}$. The size dispersion of the nanoparticles (~10%) was checked by transmission electron microscopy (TEM). A drop of viscous silicon oil was added on top of the samples to ensure homogeneity of heat diffusion. In the calculations, we assume that the refractive index around the nanoparticles was $n_m$ = 1.5.

Prior to spectra acquisitions, single nanoparticles were first located by imaging a 10×10 $µm^2$ area of the sample at fixed heating wavelength. For each single nanoparticle, spectra were then acquired by automatically sweeping the laser photon energy by steps of 10 *meV* across the resonance with a 2 *s* integration time per point. The photothermal signals were corrected for the wavelength dependence of the diffraction-limited laser spot size. The heating power was measured during the acquisitions and used to normalize the spectra.

Figure 2 shows examples of absorption spectra recorded for 4 individual nanoparticles of different sizes. One can already notice that both a blue-shift and a broadening of the SPR with decreasing particle sizes are clearly visible. The SPR spectra are asymmetric due to interband transitions. Indeed, in gold, the energy threshold for interband transitions lies at ~2.4 *eV* and is preceded by an absorption tail starting at about 1.8 *eV* [1]. Defining a full-width-at-half maximum of absorption spectra for gold nanoparticles would thus be delicate, and even impossible for very small particles. As a consequence, we extracted from each spectrum the peak resonant energy $E_R$, and the homogenous red half-width at half maximum $\Gamma_{1/2} = E_R - E_{1/2}$ where $E_{1/2}$ is the photon energy for which the absorption is half of the peak value (see Figure 2a)[19, 20].

For each particle size, we built the histograms of $E_R$ and $\Gamma_{1/2}$ (Figure 3a to h) from more than 30 single particle SPR spectra. The distributions are unimodal, but fairly wide and will be used in the following to precise the overall features observed in Figure 2. We first compare in Figure 4a the average values of $E_R$ deduced from the histograms plotted in Figure 3a to d with the size dependence predicted by Mie theory. The red-shift experienced by $E_R$ for increasing sizes is well reproduced by the theory (Figure 4a, dotted line) using the bulk dielectric function of gold[21]. Concerning the dispersion of the measured values $E_R$ (standard deviations of 17, 20, 14, 18 *meV* for particles diameters of 33, 20, 10, 5 *nm* respectively), it cannot be explained by the dispersion in the nanoparticles diameters. On the contrary, a slight ellipticity of the particles which are randomly oriented with respect to the laser polarization can explain this dispersion. Indeed, Figure 3i to l reports the distributions of the particles aspect ratios deduced from TEM images. All distributions are narrow with a mean aspect ratio of 1.1 ± 0.1, and by extending the Mie theory to the case of these randomly oriented non-spherical particles[8, 22], we predict a dispersion in $E_R$ of ~20 meV. As a consequence, although variations of $E_R$ due to inhomogeneities of the dielectric environment cannot be completely ruled out, we find that the measured width of $E_R$ can fully be explained by the ellipticity of the nanoparticles.

Mie theory with the bulk dielectric functions only predicts a minor dependence of the SPR linewidth with particle size (dashed line in Figure 4b). In order to explain the experimental variations of $\Gamma_{1/2}$ with $D$ (scattered dots in Figure 4b), surface effects have to be considered through a size-dependent dielectric function[1]:

$$\varepsilon(\omega, D) = \varepsilon_{IB}(\omega) + 1 - \frac{\Omega_p^2}{\omega^2 + i\omega\gamma(D)}$$

where $\Omega_p$ is the bulk plasma frequency. $\varepsilon(\omega, D)$ contains two contributions: the first, $\varepsilon_{IB}(\omega)$, is due to interband transitions and assumed to be size independent, whereas the second is the Drude-Sommerfeld free electron term. It contains the size-dependent phenomenological damping constant:

$$\gamma(D) = \gamma_0 + 2\frac{Av_F}{D}$$

where $\gamma_0$ is the bulk damping rate and $v_f$ is the Fermi velocity of the electrons. For gold, $\hbar\Omega_p = 9.0$ eV, $\hbar\gamma_0 \approx 70$ meV and $v_f = 1.4$ nm fs$^{-1}$. The dimensionless size parameter $A$ accounts for additional surface damping terms. More precisely, surface scattering is usually described in terms of inelastic collisions which shorten the electron mean free path when the nanoparticle sizes decrease. Additionally, for embedded nanoparticles, chemical interface damping, i.e. fast energy transfer between the nanoparticle and its close surrounding, may also damp the phase coherence of the collective oscillation[1,14] when the particle plasmon energy is close to adsorbate energy levels. As the number of conduction electrons participating in the collective oscillation is proportional to the volume of the nanoparticle, the size dependent correction in $\gamma(D)$ scales as the surface-to-volume ratio i.e. $\frac{1}{D}$.

Using the Mie theory with the size-dependent dielectric function, we simulated the absorption spectra for the different particles sizes (solid lines in Figure 2) and found $A = 0.25$ as the best value for the size parameter. From those simulations, we could then extract a theoretical red half-width at half maximum $\Gamma_{1/2}$. One should mention that $\Gamma_{1/2}$ contains the two damping contributions mentioned above. We now compare in Figure 4, the size-dependence predicted by the Mie theory using $A = 0.25$ (the gray areas illustrate the influence of the uncertainties in the tabulated values of the bulk dielectric function[21]) with the measured average values of $E_R$ and $\Gamma_{1/2}$. First, Figure 4(a) shows that introduction of the non-zero value of $A$ does not significantly affect the expected values of $E_R$ which maintains the agreement between the measurements and Mie simulations. Second, the theoretical and experimental variations of $\Gamma_{1/2}$ with $D$ are shown in Figure 4(b). We find for the same value of $A$, a good agreement between the experimental widths and Mie simulations, a clear signature of the presence of intrinsic size effects in the absorption spectra of small nanoparticles.

Interestingly, we observe notable dispersions in the measured values of $\Gamma_{1/2}$ which are more pronounced for the smallest particles (standard deviations of 13, 15, 20, 25 meV for particles diameters of 33, 20, 10, 5 nm respectively). Nanoparticle ellipticity can lead to measurements with significantly broadened SPR lines[22]. However, since the aspect ratio of our particles is 1.1 (Figure 3i to l), we calculate that the ellipticity contributes for ~10 *meV* in the dispersion of the $\Gamma_{1/2}$ histograms. Therefore it accounts only partially for the measured dispersions. Heterogeneities in the interface decay channels[1], for which the smallest nanoparticles are most sensitive can explain the additional dispersion.

For gold nanoparticles, because of the non-Lorentzian profile of the extinction cross section, the bandwidth usually used to define the plasmon dephasing time is the red half width of the absorption band[19, 20]: $T_2 = \hbar/\Gamma_{1/2}$. We found values of $T_2$ ranging from 5.9 *fs* for 33 *nm* nanoparticles down to 4.1 *fs* for 5 *nm* nanoparticles. These values are larger than those deduced from ensemble measurements[19]. This difference can be attributed to inhomogeneous broadenings in ensemble measurements.

In conclusion, we have measured for the first time the absorption spectra of individual gold nanoparticules as small as 5 *nm*. We have clearly evidenced intrinsic size effect in the optical response of the smallest particles and accessed the homogeneous width of their SPR. Future developments include the study of silver nanoparticles for which interband damping plays minor role in the SPR bandwidth and the spectroscopy of complex structures made of small nanoparticles.

ACKNOWLEDGMENT: We thank O. Lambert and A. Brisson for the TEM measurements, V. Sandoghdar for valuable discussions, O. Labeau and G. A. Blab for their assistance and P. Morin (Coherent France) for the loan of the dye laser. This research was funded by the CNRS and the MENRT (ACI Nanoscience and DRAB) and by Région Aquitaine.

FIGURE CAPTIONS

**Figure 1:** (a) Schematic of the experimental setup. The heating beam intensity is stabilized by an electro-optic modulator. It is overlaid with the probe beam on a broad band dichroic mirror. A combination of a polarizing cube and quarter wave plate is used to extract the reflected probe beam which contains the photothermal signal. (b) and (c) are 1 $\mu m^2$ photothermal images of a single 5 *nm* gold nanoparticle, taken at two different heating beam photon energy of 2.41 *eV* and 2.15 *eV* respectively. The heating power was less than 500 *µW* and the integration time per pixel was 10 *ms*.

**Figure 2:** Normalized absorption spectra of single nanoparticles of various diameters. (a) 33 *nm*, (b) 20 *nm*, (c) 10 *nm*, and (d) 5 *nm*. An example of the extracted red width at half maximum $\Gamma_{1/2}$ is shown in (a). The experimental values (open circles) are compared with simulations based on Mie theory (solid lines) using a size parameter of A = 0.25.

**Figure 3:** Histograms of $E_R$ ((a)-(d)) and $\Gamma_{1/2}$ ((e)-(h)) extracted from single particle SPR spectra. The average size of the studied nanoparticles is 33 *nm* ((a), (e)), 20 *nm* ((b), (f)), 10 *nm* ((c), (g)) and 5 *nm* ((d), (h)). For each particle size, the histograms are built from more than 30 single nanoparticle SPR spectra. The aspect ratios histograms of the nanoparticles (33 *nm* (i), 20 *nm* (j), 10 *nm* (k) and 5 *nm* (l)) are extracted from TEM images (insets). The disparity in the total number of particles contributing to these histograms is due to variations in TEM samples preparation.

**Figure 4:** Size dependence of $E_R$ (a) and $\Gamma_{1/2}$ (b). Experimental data (circles with standard deviations) are compared with Mie theory for A = 0 (dotted line) and A = 0.25 (gray area). The gray area accounts for the experimental uncertainties on the bulk dielectric function of gold given in [21].

REFERENCES

1. Kreibig, U.; Vollmer, M., *Optical properties of metal clusters*. ed.; Springer-Verlag: Berlin, **1995**.
2. Moskovits, M. *Review of Modern Physics* **1985,** 57, 783.
3. Nie, S.; Emory, S. R. *Science* **1997,** 275, (5303), 1102-6.
4. Sokolov, K.; Chumanov, G.; Cotton, T. M. *Analytical Chemistry* **1998,** 70, (18), 3898-3905.
5. Ditlbacher, H.; Krenn, J. R.; Schider, G.; Leitner, A.; Aussenegg, F. R. *Applied Physics Letters* **2002,** 81, (10), 1762.
6. Taton, T. A.; Mirkin, C. A.; Letsinger, R. L. *Science* **2000,** 289, (5485), 1757-60.
7. Schultz, S.; Smith, D. R.; Mock, J. J.; Schultz, D. A. *Proc Natl Acad Sci U S A* **2000,** 97, (3), 996-1001.
8. Bohren, C. F.; Huffman, D. R. *J. Wiley (New York)* **1983**.
9. Mie, G. *Annalen der Physik (Leipzig)* **1908,** 25, 377.
10. Link, S.; El Sayed, M. A. *Int. Reviews in Physical Chemistry* **2000,** 19, (3), 409-453.
11. Sönnichsen, C.; Geier, S.; Hecker, N. E.; von Plessen, G.; Feldmann, J.; Ditlbacher, H.; Lamprecht, B.; Krenn, J. R.; Aussenegg, F. R.; Chan, V. Z.-H.; Spatz, J. P.; Möller, M. *Applied Physics Letters* **2000,** 77, (19), 2949-2951.
12. Mock, J. J.; Barbic, M.; Smith, D. R.; Schultz, D. A.; Schultz, S. *Journal of Chemical Physics* **2002,** 116, (15), 6755-6759.
13. Sonnichsen, C.; Franzl, T.; Wilk, T.; von Plessen, G.; Feldmann, J.; Wilson, O.; Mulvaney, P. *Physical Review Letters* **2002,** 88, (7), 077402.
14. Bosbach, J.; Hendrich, C.; Stietz, F.; Vartanyan, T.; Trager, F. *Physical Review Letters* **2002,** 89, (25), 257404.
15. Lindfors, K.; Kalkbrenner, T.; Stoller, P.; Sandoghdar, V. *Physical Review Letters* **2004,** 93, (3), 037401.
16. Boyer, D.; Tamarat, P.; Maali, A.; Lounis, B.; Orrit, M. *Science* **2002,** 297, (5584), 1160-1163.
17. Berciaud, S.; Cognet, L.; Blab, G. A.; Lounis, B. *Physical Review Letters* **2004,** 93, (25), 257402.
18. Link, S.; El-Sayed, M. A. *Annual Review of Physical Chemistry* **2003,** 54, 331-66.
19. Link, S.; El Sayed, M. A. *Journal of Physical Chemistry B* **1999,** 103, 4212-4217.
20. Pinchuk, A.; von Plessen, G.; Kreibig, U. *Journal of Physics D: Applied Physics* **2004,** 37, 3133-3139.
21. Johnson, P. B.; Christy, R. W. *Physical Review B* **1972,** 6, (12), 4370.
22. Kalkbrenner, T.; Hakanson, U.; Sandoghdar, V. *Nano Letters* **2004,** 4, (12), 2309-2314.

Figure 1:

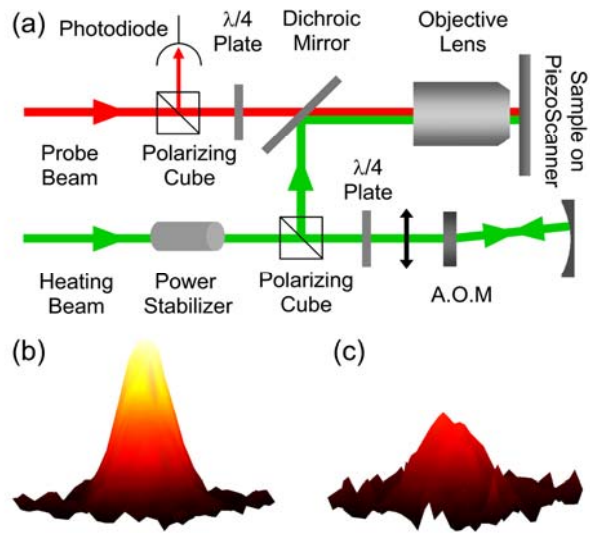

Figure 2:

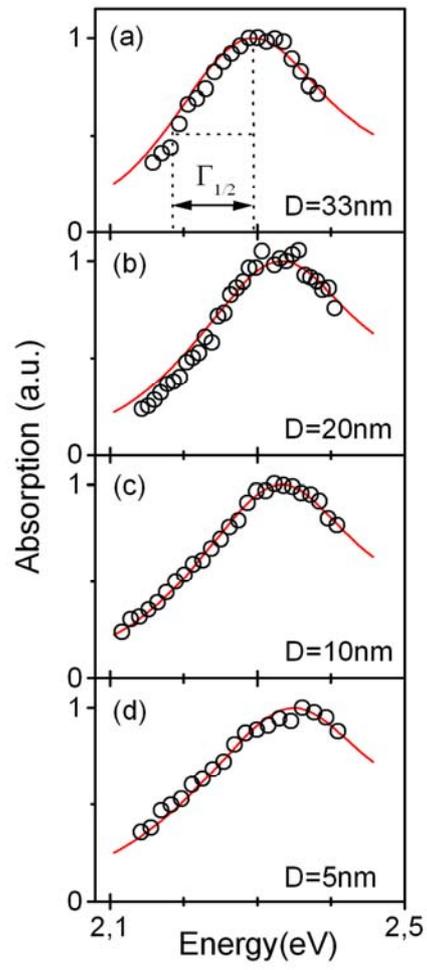

Figure 3:

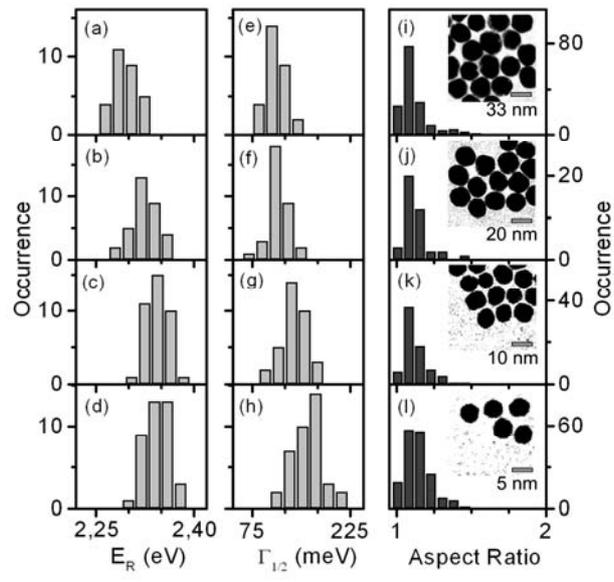

Figure 4:

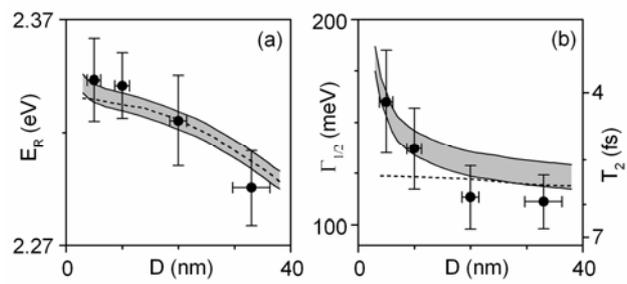